\documentclass{aa}
\usepackage{psfig}
\def\solar {\ifmmode_{\mathord\odot} \else $_{\mathord\odot}$\fi}
\def\Msol {\ifmmode {\,{\it M}\solar} \else $\,M$\solar\fi}     
\def\Rsol {\ifmmode {\,{\it R}\solar} \else $\,R$\solar\fi}     
\def\Lsol {\ifmmode {\,{\it L}\solar} \else $\,L$\solar\fi}     

\begin{document}
\thesaurus{06     
              (08.02.3;  
              08.02.4;  
              08.02.6;  
              08.12.2  
              03.20.7;  
              08.12.1)  
    }
\title{Accurate masses of very low mass stars:\\
  II The very low mass triple system Gl~866 
 \thanks{Partly based on observations made at the Observatoire de Haute 
   Provence (CNRS), and at the CFH Telescope, operated by the NRCC, the CNRS
   and the University of Hawaii}
 }
\authorrunning{Delfosse et al.}
\titlerunning{The very low mass triple system Gl~866} 
\author{X.~Delfosse \inst{1}, T.~Forveille \inst{2}, S. Udry \inst{3}, 
J.-L Beuzit \inst{2,4}, M.~Mayor \inst{3} and C.~Perrier \inst{2}     
}
\offprints{Xavier Delfosse, e-mail: delfosse@ll.iac.es}
\institute{   Instituto de Astrofisica de Canarias
              E-38200 La Laguna, Tenerife, 
              Canary Islands, Spain
\and
              Observatoire de Grenoble,
              414 rue de la Piscine,
              Domaine Universitaire de S$^{\mathrm t}$ Martin d'H\`eres,
              F-38041 Grenoble,
              France
\and
              Observatoire de Gen\`eve,
              51 Ch des Maillettes,
              1290 Sauverny,
              Switzerland   
\and
              Canada-France-Hawaii Telescope Corporation, 
              P.O. Box 1597,
              Kamuela, HI 96743, 
              U.S.A.
}
\date{Received ; Accepted}
\maketitle
   \begin{abstract}

We present very accurate orbital parameters and mass measurements
(2.4\% accuracy) for the well known very low mass triple system
Gl~866. We obtain first orbital elements for the short-period orbit
and greatly improve the long period orbit. All three stars have masses
close to 0.1 \Msol, and the system thus provides the strongest
constraints to date on the mass-luminosity relation close to the brown
dwarf limit.

   \end{abstract}

\section{Introduction}
The nearby multiple system Gl~866 has attracted considerable attention
over the last decade, as it comprises some of the coolest dwarfs for
which dynamic masses have been measured.  At d=3.45~pc (Van Altena et
al. 1995) the Gl~866 system ($\alpha$=22:38:34; $\delta$=-15:18:02;
eq=2000.0) is a very close neighbour of ours, with an
M5.5V joint spectral type (Henry et al. 1994). Leinert et al. (1986) and
McCarthy et al. (1987) independently discovered it to be a binary from
IR 1-D speckle observations, as well as Blazit et al. (1987) from
visible speckle observations. Leinert et al.  (1990) first determined
the elements of the orbit (P=2.2 years) from three years of speckle
observations, and Heintz (1993) used astrometric measurements to
improve the period to P=2.27 years.  Leinert et al. (1990) derived an
accurate ($\sim$5\%) total mass for the system, which however was too
large for the measured luminosities. For the last few years, and at
least amongst very low mass star observers, it has been common
knowledge that the brighter component of the 0.35'' pair is itself a
short period binary (Gl~866~AC), though 
the paternity of this discovery is unclear and, to our knowledge, 
any paper claim it.
This resolves the mass discrepancy identified by Leinert et al. (1990).

In this letter we present new radial velocity and angular separation
measurements, which we use to determine substantially improved
elements for the outer orbit, together with first orbital elements for
the inner one. We then proceed to briefly discuss the mass-luminosity
relation for the bottom of the main sequence, in the light of the
derived masses.

\section{Observations and data analysis}
\begin{figure} \psfig{height=5cm,file=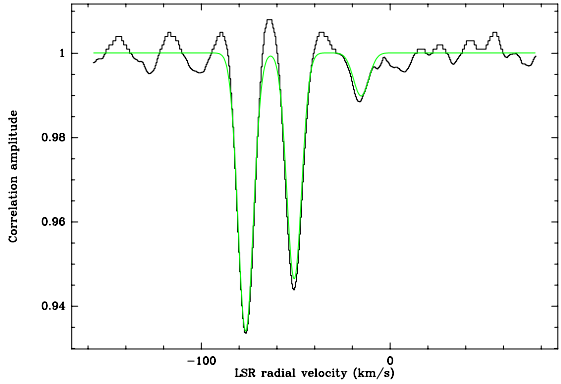,angle=-90}
\caption{The ELODIE correlation profile on julian day 2550389.3191 
  (a well separated configuration of the three peaks), and its
  adjustment by ORBIT. }
\label{correlation}
\end{figure}

Radial velocity measurements were obtained with the ELODIE
spectrograph (Baranne et al. 1996) on the 1.93m telescope of the
Observatoire de Haute Provence (France) between September 1995 and
July 1999. This fixed configuration dual-fiber-fed echelle
spectrograph covers in a single exposure the 390-680~nm spectral
range, at an average resolving power of 42000. The spectra were
analysed by numerical cross-correlation with an M4V binary (i.e. 0/1)
mask, as described by Delfosse et al. (1999a). With this setup Gl~866
is a clear triple-lined system, with relative depths for the three
well separated correlation peaks of $\sim$7, $\sim$6 and $\sim$1\% 
(Figure \ref{correlation}). 
The individual radial velocities have typical accuracies of
50-80~$\rm{m.s^{-1}}$ for the two brighter components and
200-500~$\rm{m.s^{-1}}$ for the fainter one.

Our orbital analysis largely relies on the numerous speckle
measurements of the Gl~866 system obtained by Leinert et
al. (1990). We have also obtained one angular separation measurement
at the 3.6-m Canada-France-Hawaii Telescope (CFHT) with the Adaptive
Optics Bonnette adaptive system PUE'O (Rigaut et al. 1998) and the KIR
infrared camera (Doyon et al. 1998): ${\rho}=.200{\pm}.001''$ and
${\theta}=204.1{\pm}0.9$ on July 23$^{rd}$ 1998,
Delfosse et al. (1999a) provide a detailed
description of the observing and analysis procedure, which we don't
repeat here. As the maximum separation between components A and C is
$\sim$0.010'' (Section 3), all separation observations to date,
whether Speckle, Adaptive Optics or with HST (Henry et al.  1999),
only resolve the AC photocenter from the B components.

We have used the ORBIT program (Forveille et al. 1999) to determine
the elements of both orbits through a least square adjustment to all
available observations, including the trigonometric parallax of
0.2943''$\pm$0.0035'' (Van Altena et al. 1995). ORBIT directly
supports triple systems (as well as symetrical quadruples), as long as
three-body effects can be neglected. We have not used the individual
velocities
but instead made use of a recent improvement to ORBIT (see Forveille et
al. 1999 for details) to directly adjust the orbits to the cross-correlation
profiles. 
This direct adjustment to the profiles significantly improves the
accuracy of the orbital parameters, since, for N spectroscopic
measurements, it effectively removes $\sim$2$\times$(N-1) free
parameters per spectroscopic component.  This gain is particularly
important for Gl~866, whose three correlation peaks blend for many
velocity configurations, and whose weaker C component is sometimes
only detected with a moderate signal-to-noise ratio.

In this process we produce directly a final orbit and not a intermediate 
radial velocity and we can not make available they, but we publish the 
individual radial velocity (available in the electronic form of this paper)
obtained by classical Gaussian fits of the correlation dips. We plan
to send directly (by e-mail) our profile corelation at all 
people interested to work directly with them. 

\section{Orbitals elements and masses of Gl~866}
\begin{table}
\begin{tabular}{|lll|} \hline
\multicolumn{3}{|c|}{Masses}\\ \hline
M$_{\rm{A}}$ (in \Msol) & 0.1216 & $\pm$~0.0029  \\
M$_{\rm{B}}$ (in \Msol)& 0.1161 & $\pm$~0.0029  \\
M$_{\rm{C}}$ (in \Msol)& 0.0957 & $\pm~$0.0023  \\ \hline \hline
 & Outer Orbit (AC-B) & Inner Orbit (A-C) \\ \hline
a1 (in a.u.)& 0.775 & 0.012 \\
a2 (in a.u.)& 0.414 & 0.016 \\ 
orbital parallaxe & 290.4 & $\pm$~2.7 mas\\ \hline 
\end{tabular}
\caption{Derived physical parameters.}
\label{mass}
\end{table}

\begin{figure}
\tabcolsep 0.2cm
{\begin{tabular}{cc}
{\hspace{0.1cm}
\begin{minipage}[b]{1.7cm}
\caption{Visual outer orbit. 
\vspace{1cm}}
\end{minipage}} &
\psfig{height=5cm,file=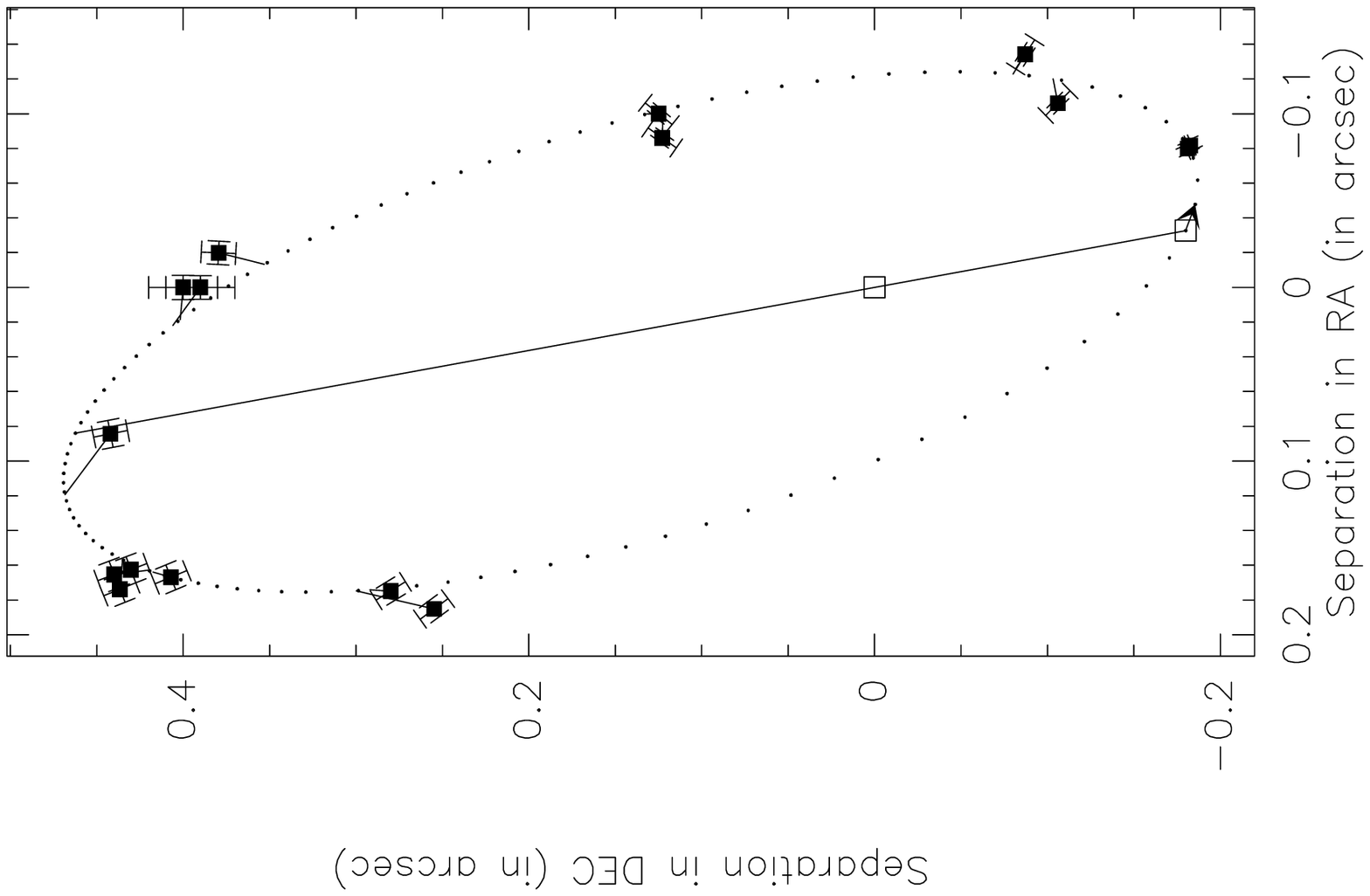,angle=-90} \\
\end{tabular}}
\label{orbit2}
\end{figure}

\begin{figure*}
\tabcolsep 0.5cm
{\begin{tabular}{cc}
\psfig{height=5.8cm,file=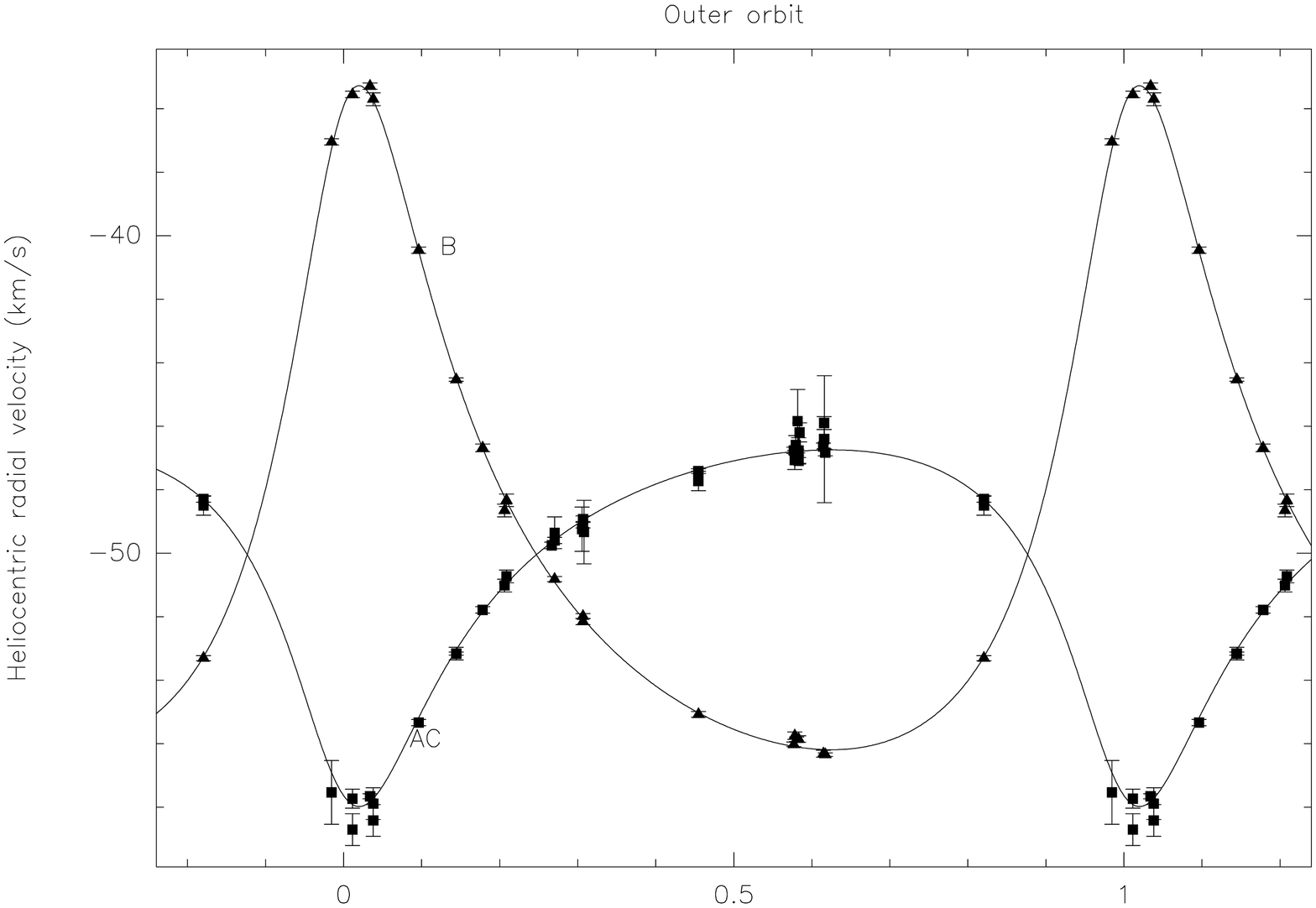,angle=-90} &
\psfig{height=5.8cm,file=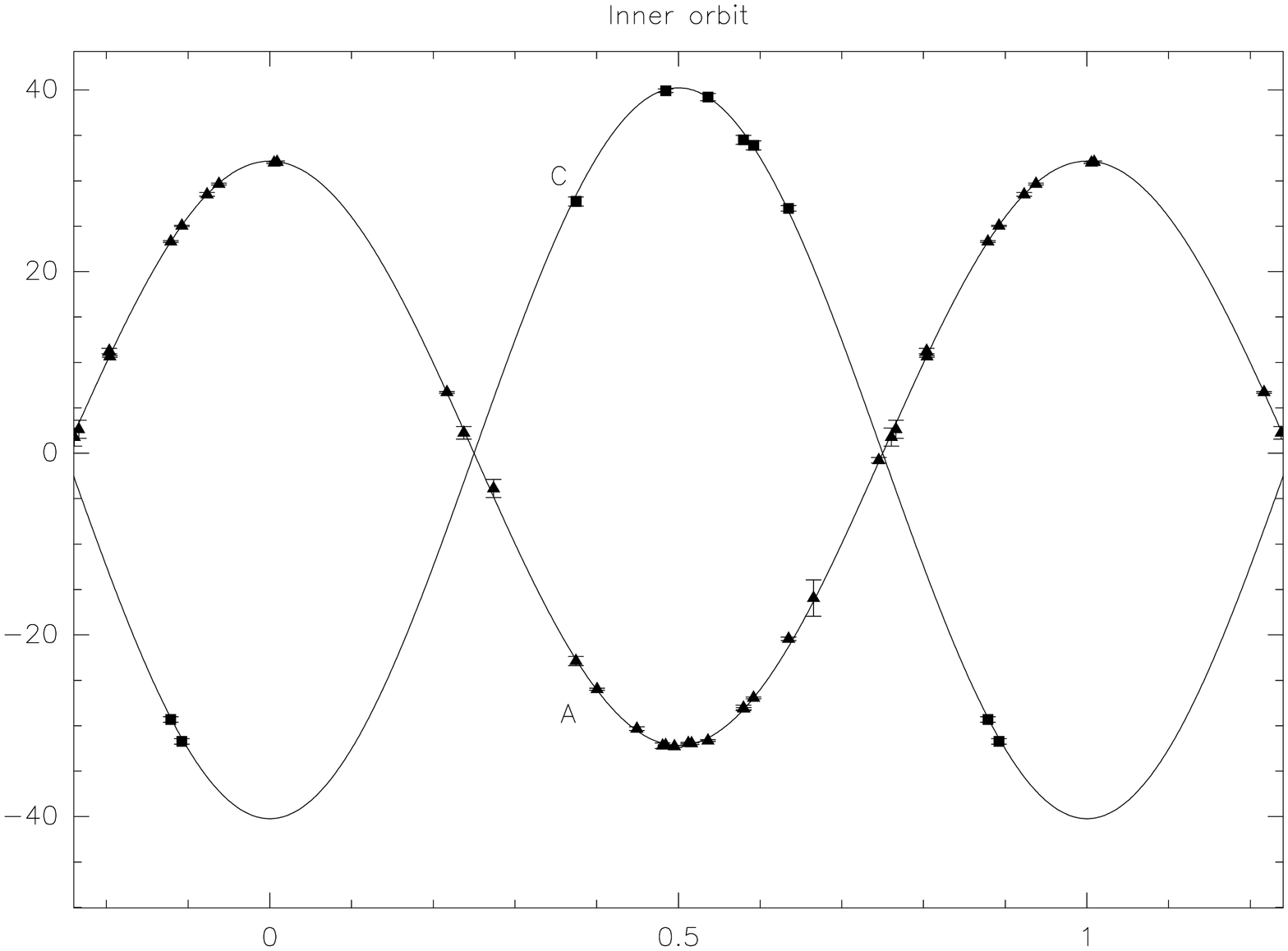,angle=-90} \\
\end{tabular}}
\caption{Phased radial velocity curves for the outer and inner 
orbits of Gl866. In both cases the contribution of the other orbit has 
been subtracted from the displayed measurements.}
\label{orbit}
\end{figure*}

\begin{table*}
\begin{tabular}{llllllllll} \hline
& \multicolumn{1}{c}{P} & \multicolumn{1}{c}{T$_0$} & \multicolumn{1}{c}{e} & 
\multicolumn{1}{c}{a} & \multicolumn{1}{c}{i} & \multicolumn{1}{c}{$\Omega$}
& \multicolumn{1}{c}{$\omega$} & \multicolumn{1}{c}{K$_1$} &
\multicolumn{1}{c}{K$_2$} \\
& \multicolumn{1}{c}{(days)} & \multicolumn{1}{c}{(Julian day)}  & & 
\multicolumn{1}{c}{(arcsec)} & \multicolumn{1}{c}{(deg)} & 
\multicolumn{1}{c}{(deg)} & \multicolumn{1}{c}{(deg)} & 
\multicolumn{1}{c}{(${\rm km\,s^{-1}}$)} &
\multicolumn{1}{c}{(${\rm km\,s^{-1}}$)} \\ \hline
Leinert et al. (1990) & 803.5 & & 0.41 & 0.36 & 112 & 163& 340 & & \\
(Outer Orbit (AC-B)) & $\pm$~8 & & $\pm$~0.01 & $\pm$~0.01 & $\pm$~2 & $\pm$~2 & $\pm$~5 & & \\
This letter & 822.6 & 48517 & 0.446 & 0.346 & 112 & 161.5 & 337.6 &
10.62 &  5.64 \\
(Outer Orbit (AC-B)) & $\pm$~0.6 & $\pm$~2.5 &  $\pm$~0.003 & $\pm$~0.002 & $\pm$~1 & $\pm$~0.6 & 
$\pm$~0.6 & $\pm$~0.06 & $\pm$~0.04 \\ \hline
This letter & 3.78652 & 50640.775 & 
0.000 & &  $\sim$ 117 & & & 32.11 & 40.8 \\
(Inner Orbit (A-C)) & $\pm$~0.00001 & $\pm$~0.001 & $\pm$~0.001 & & & & & $\pm$~0.02 
& $\pm$~0.1 \\ \hline
\end{tabular}
\caption{Orbital elements of the outer (AC-B) and inner (A-C) orbits of
Gl~866.}
\label{orb_param}
\end{table*}

Figure \ref{orbit} 
and Table \ref{orb_param} 
together summarize the
orbital solution which results from this simultaneous adjustment of
both orbits to the radial velocity, angular separation, and parallax
data. 
The longitude of the periastron is given with the spectroscopic 
convention, and thus refers to the primary
%
The brightest and most massive component of the system (A) and
its faintest and lightest one (C) constitute a close binary with a
$\sim$~3.8 day period, orbiting with the third star in a 823 day
orbit.  With ${\mathrm P}_{out}/{\mathrm P}_{in}{\sim}$200, the system
is strongly hierarchical. This justifies our use of two
non-interacting keplerian orbits in the adjustment, which amounts to
neglecting all second order gravitational effects. As just an example,
the expected nodal precession period for the parameters of Gl~866,
is $T_{prec}\sim900$~years (Harrington
(1968), Mazeh and Saham (1976)), 
very much longer than the 4~years span of the radial velocity
observations (we argue below that the system is probably coplanar, so
that nodal precession in particular probably has a small or null amplitude, in
addition to this long period). We find no secular trend in the
residuals, also vindicating the use of two keplerian orbits.
Most orbital elements are essentially constrained by the spectroscopic
data alone, except for the semi-major axis, the inclination ($i$) and 
the orientation of the nodal line on sky ($\Omega$), which are only
constrained by the angular separation data and the parallax.

The inner orbit has negligible eccentricity, as expected from tidal
circularisation for its short period. The physical separation between
A and C is only 0.03 a.u. (Table \ref{mass}), but, thanks to the small
distance to Gl~866, their apparent separation is
$\sim$0.01''. Infrared interferometers with baselines longer than
$\sim$40m (such as the future VLTI) will be able to resolve this pair,
which is also an excellent target for observations as an astrometric
binary: the large mass ratio and $\sim$2~mag V band magnitude
difference combine to produce a ${\sim}{\pm}$0.005'' astrometric
amplitude in the passband of the FGS astrometers on {\it HST}, well
within the range of these instruments, and the outer B component
provides an excellent local reference.

For the time being we cannot directly determine the inclination of
the inner orbit $i_s$, but it is strongly constrained by the requirement
that M$_{\rm{A}}+$M$_{\rm{C}}$ derived from the outer orbit must
match the sum of the spectroscopic M${\times}sin^3{i_s}$ obtained from 
the inner orbit. This gives $sin^3{i_s}$=0.705$\pm$0.016, and therefore
$i_s$=117$^o\pm$1 or $i_s$=63$^o\pm$1. One of those two determinations
is very close to the inclination of the outer orbit ($i_s-i_l~{\sim}~5^o$).
This probably points to a coplanar system, in keeping with a general 
tendency of close triple systems (Fekel 1981). One would however
need to determine ${\Omega}_s$ and to resolve the $i_s$ ambiguity to 
ascertain this. 

As for Gl~570BC (Forveille et al. 1999), the combination of high
accuracy radial velocity {\it and} angular separation data determines
enough orbital parameters to derive the mass of each star with very
good accuracy (Table \ref{mass}), in spite of significant
uncertainties on the elements constrained by the visual data alone:
the system is fortunately close to edge-on, so that the relatively
large uncertainty on $i$ does not unduly propagate to ${\sin}^{3}i$,
and the derived masses are fairly accurate. We obtain relative
precisions of 2.4\% for both B and the AC barycenter. The short period
spectroscopic orbit determines the mass ratio within AC
very well, so that the individual masses of A and C also have this
same precision. Amongst very low mass stars, only four systems have
more accurate mass determinations (the three detached M-dwarfs eclipsing
binaries CM~Dra, YY~Gem and GJ~2069A; and Gl~570BC). The three 
components of Gl~866 are at least
twice less massive than any of those. Gl~866A and B are slightly more 
massive than 0.1\Msol, and Gl~866C is slightly less massive.

\section{Mass-luminosity relation}
Henry et al. (1999) determine absolute magnitudes of
M$_{\rm{V}}=15.58{\pm}0.07$ for Gl~866B and (implicitly)
M$_{\rm{V}}=15.18{\pm}0.1$ for the AC pair, which
is too close for {\it HST} to resolve. The relative areas of the
cross-correlation ``dips'' in double/triple-lined multiple systems
fortunately provides a welcome handle on the luminosity ratios in such
unresolved systems. These areas are to first order proportional 
to the relative luminosities, with weaker dependences on effective
temperature and metallicity (Mayor 1985), and [Fe/H] must here be 
identical for the three components, which formed from the same 
interstellar gas 
(except in case of binary formation by capture, wich are not 
probable here).
For ELODIE spectra correlated with the M4 mask, we have unfortunately not 
yet calibrated the spectral type (and metallicity) dependence, and as a
consequence we can only derive an approximate V-band luminosity ratio
from the raw equivalent-width ratio. We obtain L$_C$/L$_{AC}$=0.158, which 
is consistent with the $\sim$2~magnitude difference in the V band between 
0.12\Msol{\ } and 0.095\Msol{\ } 5~Gyr Baraffe et al. (1998) models. We 
conservatively assign a factor of 1.5 uncertainty to this ratio,
which dominates the error bars on the luminosity of the C component.
The A component on the other hand sufficiently dominates the
light of the AC pair in the V band that even this quite large uncertainty
does not propagate into unduly large errors on its absolute V magnitude.
We obtain M$_{\rm{V}}(A)=15.34{\pm}0.14$ and M$_{\rm{V}}(C)=17.34{\pm}0.45$.

\begin{figure}
\psfig{height=6cm,file=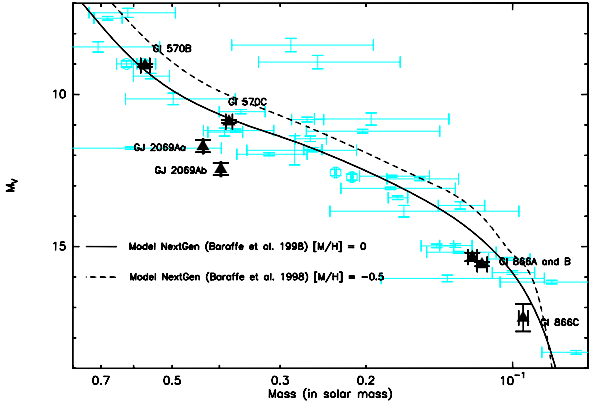,angle=-90}
\caption{V band mass-luminosity relation. The error bars without
points are data from Henry and McCarthy (1993), Henry et
al. (1999) and Torres et al. (1999). The circles represent the two well known M eclipsing
binaries, YY~Gem and CM~Dra. The triangles represent our recent
mesurements: GJ~2069A, Gl~570B and the three components of Gl~866. 
The two curves are 5~Gyr theoretical isochrones Baraffe et al. 
(1998) for two metallicities.}
\label{mass_lum}
\end{figure}

Figure \ref{mass_lum} represent our present best attempt at a V-band
mass-luminosity diagram. It shows the data points of Henry and
McCarthy (1993), Henry et al. (1999) and Torres et al. (1999), as well as our recent
measurements of Gl~866, GJ~2069Bab (Delfosse et al. 1999b),
and Gl~570BC (Forveille et al. 1999). For this last source we obtain
absolute magnitudes of (M$_{\rm{V}}=9.05{\pm}0.05$ and
M$_{\rm{V}}=10.89{\pm}0.05$) from the V band photometry compiled by
Leggett (1992), 
the HIPPARCOS parallax, and the V-band magnitude difference of
Henry et al. (1999).  The GJ~2069Bab eclipsing pair is clearly
subluminous for its mass, most likely because its metallicity is significantly
larger than solar (Delfosse et al. 1999b). The other objects are quite
well fitted by the 5~Gyr isochrones of Baraffe et al. (1998), though
the three Gl~866 component are slightly less luminous ($\sim$0.5~mag) in the V
band than the solar metallicity model for their mass. This discrepancy
most likely reflects a known low level problem in the models: the
present generation of M-dwarf atmospheric models is specifically
suspected to lack an unidentified opacity source in the V passband,
which at the effective temperature of late M dwarfs would contribute
about $\sim$0.5~mag of extra absorption in this band (Allard, private
communication). 
Alternatively, the observations of Gl~866 can be
reconciled with the models if its metallicity is slightly larger than
that of the sun. 
With the recent and forthcoming improvements in the
accuracy of M dwarfs mass measurements, metallicity determinations
will increasingly become a crucial limiting factor in accurate
comparisons with stellar models, as has long been the case for more
massive stars (e.g. Andersen 1991). Quantitative metallicity
measurements of M dwarfs are difficult in the optical range (e.g. Valenti, 
Piskunov \& Johns-Krull 1998), but K band spectroscopy probably has the
potential to reach the necessary accuracy (Allard, private
communication).

\begin{acknowledgements}
We thank the technical staffs and telescope operators of OHP and CFHT
for their support during these long-term observations. We also thank
Gilles Chabrier, Isabelle Baraffe and France Allard for many useful
discussions on very low mass star models. We are grateful to the
anonymous referee for very useful comments.
\end{acknowledgements}

\newpage

\begin{table*}
\begin{tabular}{|ll|lll|} \hline
Julian Day   &   civil dates V$_A$               &     V$_B$               &    V$_C$ \\ 
2400000+ & & (in km/s) & (in km/s) & (in km/s) \\\hline \hline
50019.325   & 28-OCT-1995 19:47   &   -25.004  $\pm$  .100  &     -53.301 $\pm$   .080  &       -77.626 $\pm$   .300 \\  
50313.543   & 18-AUG-1996 01:01  &   -79.960  $\pm$  .100  &     -46.680 $\pm$   .120  & \\
50386.336   & 29-OCT-1996 20:03  &   -38.987  $\pm$  .100  & & \\
50389.319   & 01-NOV-1996 19:39  &   -76.516  $\pm$  .100  &     -50.815 $\pm$   .080  &       -15.680 $\pm$   .500 \\
50418.269   & 30-NOV-1996 18:26  &   -46.732  $\pm$  .700  & & \\
50419.244   & 01-DEC-1996 17:51  &   -81.252  $\pm$  .100  &     -52.148 $\pm$   .100  & \\
50419.324   & 01-DEC-1996 19:46  &   -80.921  $\pm$  .100  &     -51.979 $\pm$   .080  & \\
50420.249   & 02-DEC-1996 17:59  &   -47.177  $\pm$ 1.000  & & \\
50641.595   & 12-JUL-1997 02:17  &   -40.094  $\pm$  .100  &     -56.016 $\pm$   .070  & \\
50642.594   & 13-JUL-1997 02:15  &   -78.992  $\pm$  .300  &     -55.744 $\pm$   .120  & \\
50643.596   & 14-JUL-1997 02:18  &   -47.563  $\pm$  .300  & & \\
50644.594   & 15-JUL-1997 02:16  &   -14.713  $\pm$  .100  &     -55.855 $\pm$   .070  & \\
50645.598   & 16-JUL-1997 02:21  &   -50.656  $\pm$ 1.000  & & \\
50646.592   & 17-JUL-1997 02:11  &   -78.420  $\pm$  .100  &     -55.849 $\pm$   .100  &         -7.567 $\pm$   .400 \\
50647.605   & 18-JUL-1997 02:30  &   -35.530  $\pm$  .300  & & \\
50672.583   & 12-AUG-1997 02:00  &   -72.725  $\pm$  .110  &     -56.314 $\pm$   .110  & \\
50673.471   & 12-AUG-1997 23:17  &   -67.182  $\pm$  .200  &                           &        -19.782 $\pm$   .300 \\
50674.617   & 14-AUG-1997 02:48  &   -17.102  $\pm$  .100  &     -56.341 $\pm$   .060  & \\
51017.580   & 23-JUL-1998 01:55  &   -89.724  $\pm$  .080  &     -35.289 $\pm$   .100  & \\
51108.353   & 21-OCT-1998 20:27  &   -85.172  $\pm$  .050  &     -44.530 $\pm$   .050  &        -13.116 $\pm$   .200 \\
51159.238   & 11-DEC-1998 17:43  &   -22.522  $\pm$  .200  &     -48.651 $\pm$   .200  & \\
51161.231   & 13-DEC-1998 17:32  &   -81.280  $\pm$  .200  &     -48.331 $\pm$   .200  & \\
51363.594   & 04-JUL-1999 02:16  &   -22.324  $\pm$  .070  &     -55.070 $\pm$   .090  &        -79.100 $\pm$   .310 \\ \hline
\end{tabular}
\caption{Radial velocity measurements. To be published in electronic form
only, and available at the end of this referee version.}
\label{tab_vel}
\end{table*}

\end{document}